\documentclass[a4,12pt]{iopart}
\usepackage{graphicx}

\begin{document}
\title{\bf{Effect of energy loss on azimuthal
correlation of charm and correlated charm decay
in collision of lead nuclei at $\sqrt{s}$= 2.76 A TeV}}

\title[Effect of energy loss......]{}
\author{Mohammed Younus \footnote{Email: mdyounus@vecc.gov.in} and
 Dinesh K. Srivastava \footnote{Email: dinesh@vecc.gov.in}}

\address{Variable Energy Cyclotron Center, 1/AF, Bidhan Nagar, Kolkata,
 700 064, India}

\begin{abstract}
We present the effect of energy loss of charm/anti-charm produced
in a relativistic heavy ion collision as they traverse the resulting
quark gluon plasma on the azimuthal correlation of $c\bar{c}$ and $D\overline{D}$
pairs and correlated decay of charm into leptons.
We employ an empirical model of energy loss by charm quark energy loss
and find that the consequences are easily discernible as different cuts
on their momenta are applied. We also notice a modest increase in the invariant
mass spectrum of dileptons from correlated decay as mentioned above
due to energy loss.
\end{abstract}

Key-words: Charm quark, D mesons, non-photonic electrons, QGP, relativistic heavy ion collisions,
            NLO-pQCD, correlations

PACS Indices: 14.65.Dw, 14.40.Lb, 13.20.Fc, 13.30.Ce, 12.38.Mh, 25.75.Dw, 14.40.Pq,14.65.Py

\maketitle

\section{\bf Introduction}
Heavy quarks are emerging as valuable probes for the study of
quark gluon plasma produced in relativistic heavy ion collision.
This has its origin in the large mass of heavy quarks
which lends them quite a few advantages, viz, they
are produced at a time $\approx \frac{1}{2m_Q}$
which is smaller than the formation time of quark gluon
plasma. Their large mass ensures that their production
can be calculated reliably using perturbative QCD and they may
not be deflected substantially from their path due to collisions
with quarks and gluons and due to radiation of gluons.
The drag experienced by the heavy quarks due to these collisions
and radiations however leads to medium modification of their production
which is quite similar to those for light quarks,
as leading particles~\cite{largedrag,raa}. Recent calculations which treat the so
called 'dead cone' with more care~\cite{deadcone}, also show that heavy quarks
lose energy in a manner quite similar to light quarks~\cite{dedx}.

Recently it has been pointed out that correlations of heavy
quarks (charm and anti-charm) can add a richness to these studies
by adding several features~\cite{corr}. Consider for example heavy quark
production at lowest order of pQCD. They would be produced back to back.
The two members of the correlated pair may in general lose
different amounts of energy as they may cover differing path
lengths in the plasma. However if they do not change their
directions, they would continue to remain back-to-back. Now
consider that there is a strong flow and that the heavy quarks
take part in flow~\cite{flow}. It is now possible that one of them is produced
with a transverse momentum parallel to flow velocity and its
momentum will increase, while the momentum of the other will decrease.
In fact if the radial flow velocity $v_T$ $>$ $p_{T}/E$(charm),
the charm will change its direction and the back-to-back correlation
may change to forward correlations. When, however, the flow
velocity is not collinear with the momentum, the final momenta
will be separated by 0 $<$ $\phi$ $<$ $\pi$. Thus while the
energy loss is not likely to alter the angular correlation of heavy quarks at
lowest order in pQCD, a strong elliptic flow will bring in some
interesting and rich structure, the analysis of which could throw
some light on interplay of energy loss and flow.

There is, however, a substantial production of heavy quarks at next
to leading order in pQCD. The NLO process $gg \rightarrow Q\overline{Q}$
can proceed in two ways(among others). Either one of the final state
gluons in the process $gg \rightarrow gg$ splits ($g \rightarrow Q\overline{Q}$)
or one of the heavy quarks radiates a gluon following $gg \rightarrow Q\overline{Q}$.
The pair is expected to be collinear in the first case and deviate
back-to-back in the second case. The processes where gluon is emitted
from the external legs will fill up the region 0 $<$ $\phi$ $<$ $\pi$.
Now energy loss will alter the correlations in a complex manner. If our
assumption on heavy quarks not changing direction due to
energy loss largely holds then $p_T$ integrated correlation is
likely to remain unchanged. However if we now study the correlation
for different cuts on $p_T$, some interesting patterns may emerge.
Different heavy quarks lose different momenta!

We can now discuss correlated decay of charm-anti-charm into
electron-positron pair. The invariant pair mass distribution
of electron pair obtained from decay shows interesting features.
It is seen earlier that large suppression
of heavy quark as seen through $R_{AA}$, results in
increase of D mesons as well as single electron spectrum
at low momentum by a few percent. This characteristic increase
is quite different from enhancement due to Cronin effect~\cite{cronin} and
is found to be due to large effective drag upon charm
by thermalized medium.
The invariant pair mass distribution
of electron pair obtained from decay shows similar feature from
effects due to energy loss by charm quarks~\cite{kampfer}.
The electron pairs pile up at low invariant mass region resulting
in characteristic enhancement in electron distribution.

In the following we study some of these features of the correlation
of heavy quarks with collision of lead nuclei at 2760 GeV/nucleon
as an example. The paper is organized as follows.
Sec. 2 contains formalism
for charm production cross-section from pp collisions and
lead on lead collision at LHC energy. Sec. 3 contains
an empirical model of energy loss employed to determine the
medium effect on charm correlation. Sec. 4 presents our
results and discussions on azimuthal correlation and correlated
charm decay. Finally Sec. 5 gives the summary followed by
acknowledgement and bibliography.

\section{\b Formulation}
The correlation of heavy quarks produced in $pp$ collisions
is defined as
\begin{equation}
E_1\,E_2\,\frac{d\sigma}{d^{3}p_{1}\,d^{3}p_{2}}=\frac{d\sigma}{dy_1\,dy_2\,d^{2}p_{T1}\,d^{2}p_{T2}}=C\,,
\label{corr1}
\end{equation}
where $y_1$ and $y_2$ are the rapidities of heavy quark and anti-quark and $\bf{p_{Ti}}$ are
the respective momenta.

At the leading order, the differential cross-section for the charm correlation for
proton on proton collision is given by

\begin{equation}
C_{LO}=\frac{d\sigma}{dy_1\,dy_2\,d^{2}p_{T}}\delta{(\bf{p_{T1}}+\bf{p_{T2}})}
\label{corr2}
\end{equation}

One can now calculate~\cite{corr,jamil}
\begin{eqnarray}
\frac{d\sigma_{pp}}{dy_1 dy_2 d^{2}p_{T}} &=& 2 x_{a}x_{b}\sum_{ij}
\left[f^{(a)}_{i}(x_{a},Q^{2})f_{j}^{(b)}(x_{b},Q^{2})
\frac{d\hat{\sigma}_{ij}(\hat{s},\hat{t},\hat{u})}{d\hat{t}}
\right.\nonumber\\
&+& \left.f_{j}^{(a)}(x_{a},Q^{2})f_{i}^{(b)}(x_{b},Q^{2})
\frac{d\hat{\sigma}_{ij}(\hat{s},\hat{u},\hat{t})}{d\hat{t}}\right]
/(1+\delta_{ij})~,
\label{sigma}
\end{eqnarray}
where $p_{T}$ and $y_{1,2}$ are the momenta and rapidities of
produced charm and anti-charm and
$x_{a} $ and $x_{b} $ are the fractions of the momenta carried by the partons
from their interacting parent hadrons.
These are given by
\begin{equation}
x_{a}=\frac{M_{T}}{\sqrt{s}}(e^{y_1}+e^{y_2})~;~~~~
x_{b}=\frac{M_{T}}{\sqrt{s}}(e^{-y_1}+e^{-y_2})~.
\label{x}
\end{equation}
where $M_{T}$ (= $\sqrt{m_{Q}^{2}+p_{T}^{2}}$), is the transverse mass
of the produced heavy quark.
The subscripts $i$ and $j$ denote the interacting partons, and $f_{i/j}$
are the partonic distribution functions for the nucleons.
The invariant amplitude, $\left|M\right|^2$ in differential cross-section
$d\hat{\sigma}/d\hat{t}$ is taken from ref.~\cite{invM}.

The processes included for LO calculations are:

\begin{eqnarray}
g+g \rightarrow c+\overline{c}\nonumber\\
q+\bar{q} \rightarrow c+\overline{c}~.
\label{processLO}
\end{eqnarray}

At Next-to-Leading order the subprocesses included are as follows:
\begin{eqnarray}
g+g \rightarrow c+\overline{c}+g\nonumber\\
q+\bar{q} \rightarrow c+\overline{c}+g\nonumber\\
g+q(\bar{q}) \rightarrow c+\overline{c}+q(\bar{q})~.
\label{processNLO}
\end{eqnarray}

The eq. \ref{corr1} gives the correlation of heavy quarks
from initial fusion in proton-proton collision. The
azimuthal correlation of heavy quark for Pb+Pb collision
at given impact parameter is given by

\begin{equation}
E_c\,E_{\bar{c}}\,\frac{dN_{AA}}{d^{3}p_c\,d^{3}p_{\bar{c}}}=T_{AA}E_c\,E_{\bar{c}}\,\frac{d\sigma_{pp}}{d^{3}p_c\,d^{3}p_{\bar{c}}}
\label{corr3}
\end{equation}

For lead on lead collisions at LHC, we have used $T_{AA}$ = 292 fm$^{-2}$ for b = 0 fm
We have used CTEQ5M
structure function.
 The factorization,
renormalization, and fragmentation scales are chosen as 2$\sqrt{m_c^2+p_T^2}$
and the charm quark mass, $m_c$ has been taken as 1.5 GeV.

\section{Energy Loss Formalism}
We use the empirical model for the energy loss for charm quarks
proposed in one of our earlier paper.
We perform a Monte Carlo implementation of our model
calculations and estimate the azimuthal correlation as well as correlated
decay of charm pair with charm cross-section determined using NLO-pQCD calculations.

We assume that the energy loss of heavy quarks proceeds
through the momentum loss per collision is given by,~\cite{empeloss}
\begin{equation}
 (\Delta p)_i=\alpha \, (p_i)^{\beta}~,
 \label{deltapt}
\end{equation}
so that one can write
\begin{equation}
\frac{dp}{dx}=-\frac{\Delta p}{\lambda}
\label{dpdx}
\end{equation}
where $\alpha$ and $\beta$ are parameters with best values at $\sqrt{s}$= 2760 GeV/nucleon
taken from publication by Younus et al~\cite{largedrag}
and $\lambda$ is the mean free path of the charm quark, taken as 1 fm throughout.
Thus the momentum of the charm quark after $n$ collisions will be given by
\begin{equation}
p_{n+1}=p_n-(\Delta p)_n
\end{equation}
The probability for the charm quark to have $n$ collisions,
 while covering the path length $L$ is given by
\begin{equation}
P(n,L)=\frac{(L/\lambda)^{n}}{n!}e^{-L/\lambda}.
\label{prob}
\end{equation}

So now we estimate the largest number
of collisions- $N$, which the charm quark
having momentum $p_T$ can undergo. Next we sample the number of
collisions $n$, which the charm undergoes from the distribution
\begin{equation}
p(n)=P(n,L)/\sum_{n=1}^N P(n,L)
\end{equation}
to get the final momentum of the charm(anti-charm) quark.

Next we fragment the charm quark using Peterson fragmentation
function given by $D$. We have assumed that
$D(z)$, where $z=p_D/p_c$, is identical for all the $D$-mesons,~\cite{peterson}
and
\begin{equation}
D^{(c)}_D(z)=\frac{n_D}{z[1-1/z-\epsilon_p/(1-z)]^2}~,
\label{frag}
\end{equation}
where $\epsilon_p$ is the Peterson parameter and
\begin{equation}
\int_0^1 \, dz \, D(z)=1~.
\end{equation}
We have kept it fixed at $\epsilon_p$=0.13.

Then we have included semileptonic decay of $D(\overline{D})$ mesons by
parameterizing electron distribution function taken from Ref.~\cite{altarelli}.
Finally we show our results for $dN_{c\overline{c}}/d\Delta\phi$, $dN_{D\bar{D}}/d\Delta\phi$,
$E_c\,E_{\overline{c}}dN/d^{3}p_{c}d^{3}p_{\overline{c}}$ and $dN/dM_{e^{+}e^{-}}$.

\begin{figure*}[h]
\begin{center}
\includegraphics[height=3in,width=3in,angle=270]{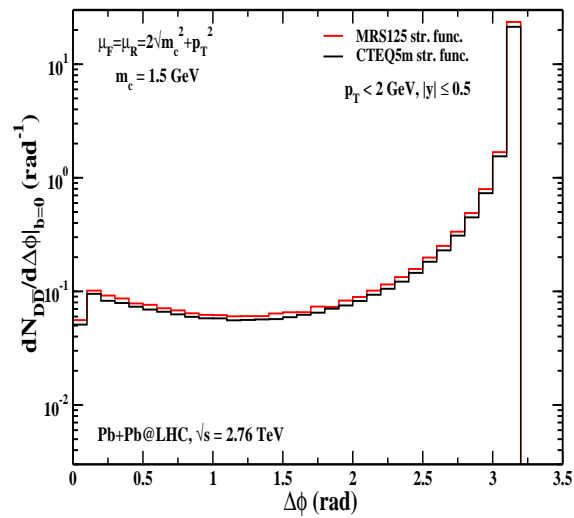}
\caption{(colour on-line)Comparison of D mesons azimuthal spectrum
for two different structure functions.}
\label{fig1}
\end{center}
\end{figure*}

\begin{figure*}[h]
\begin{center}
\includegraphics[height=3in,width=3in,angle=270]{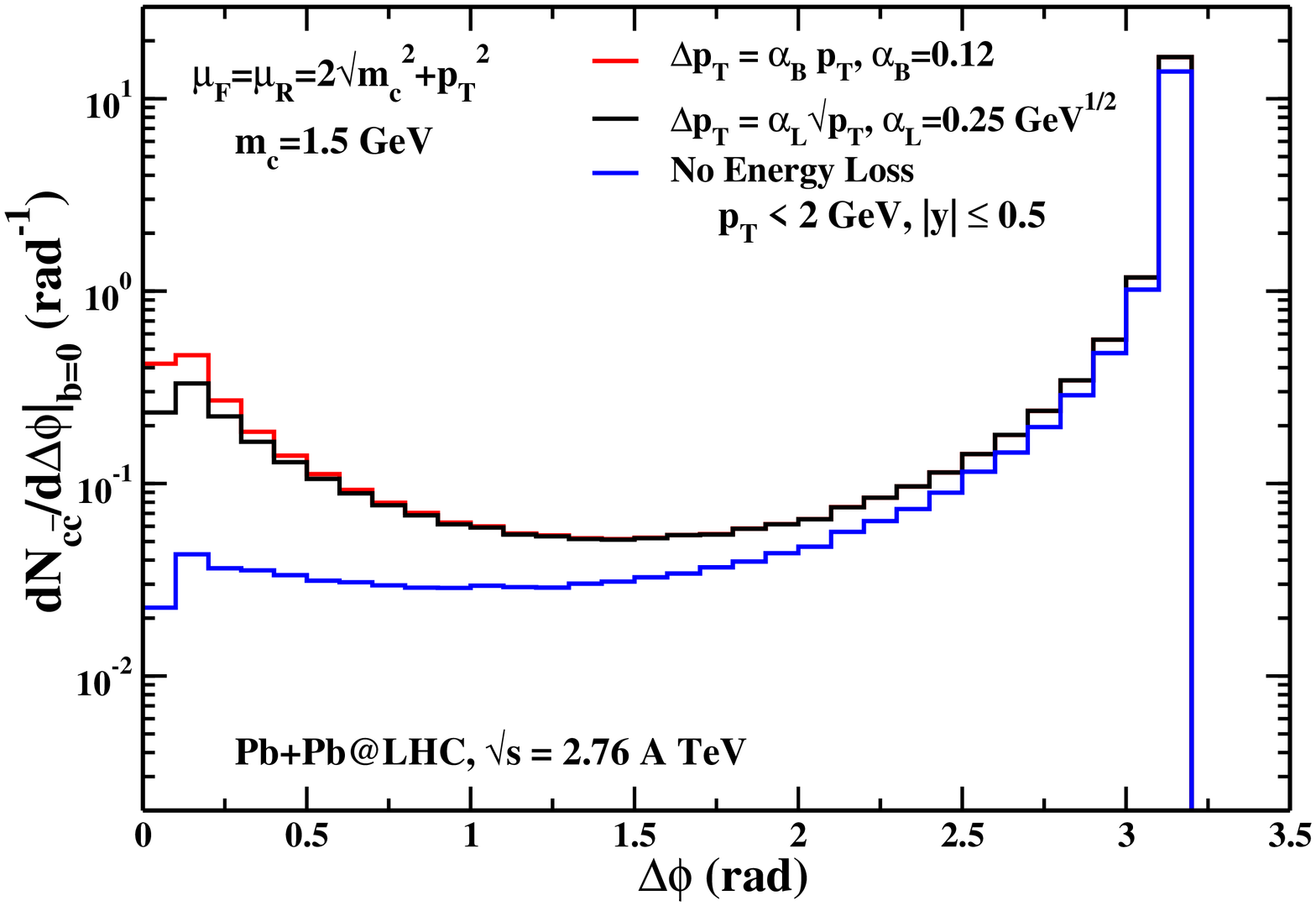}
\includegraphics[height=3in,width=3in,angle=270]{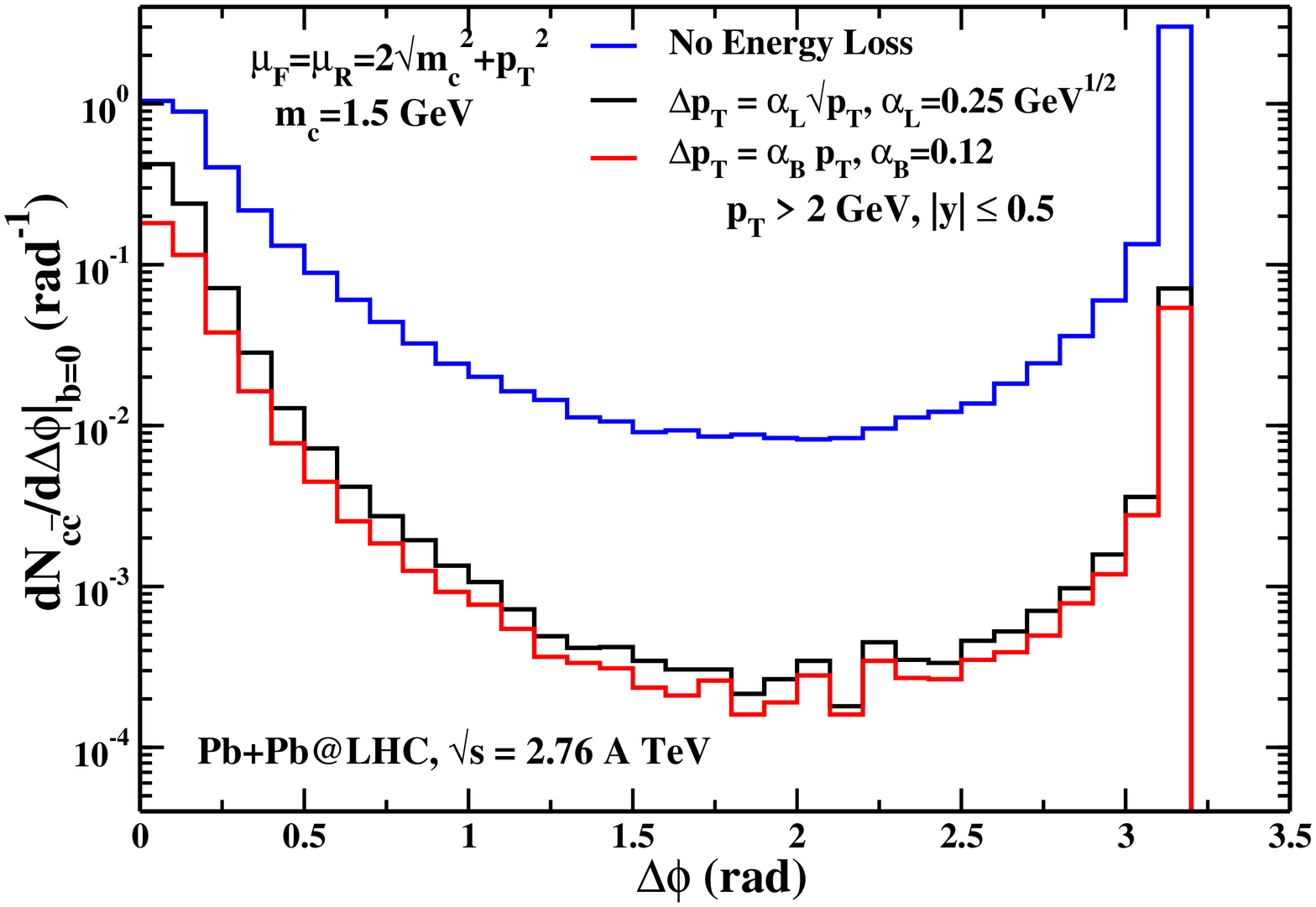}
\includegraphics[height=3in,width=3in,angle=270]{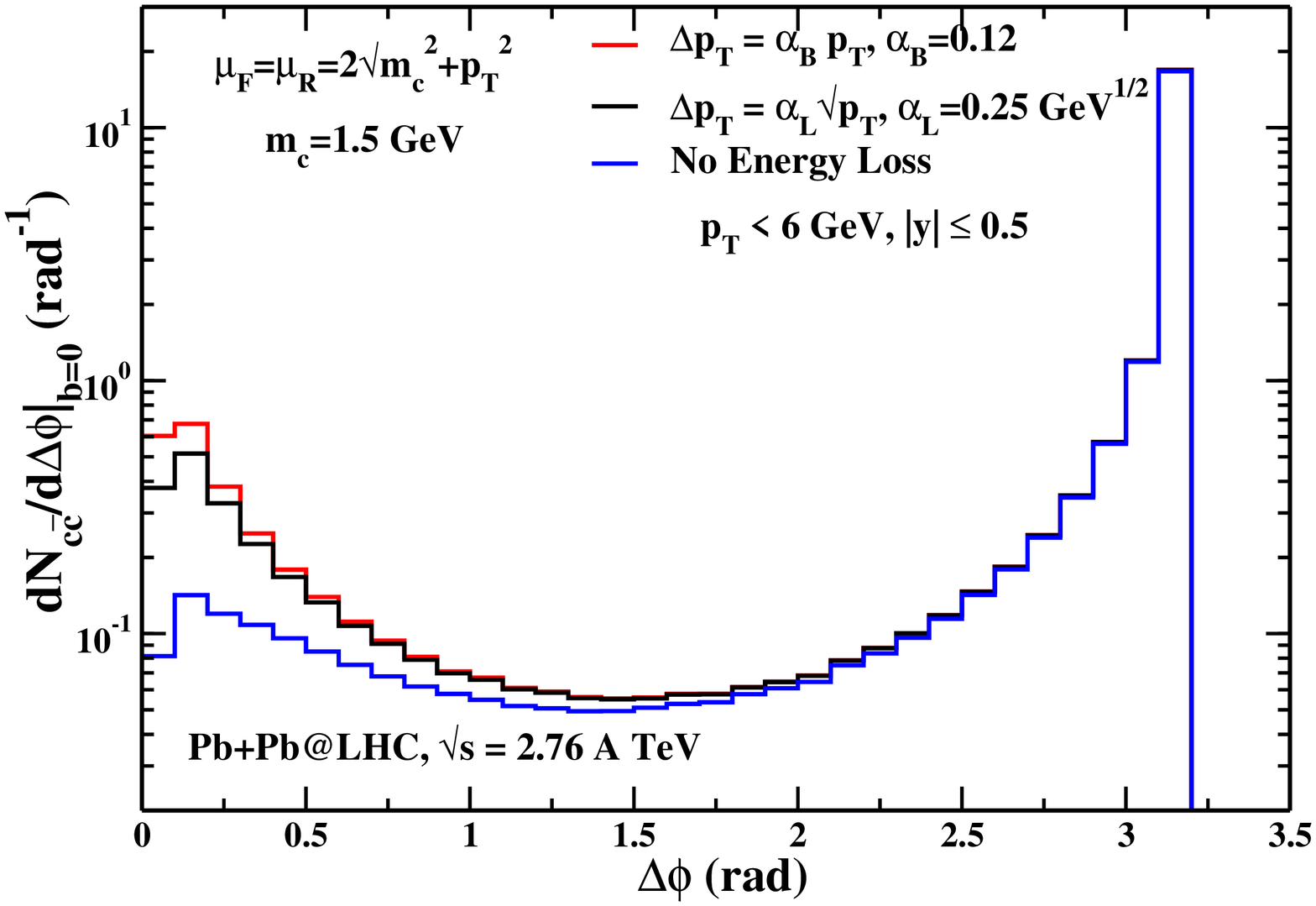}
\includegraphics[height=3in,width=3in,angle=270]{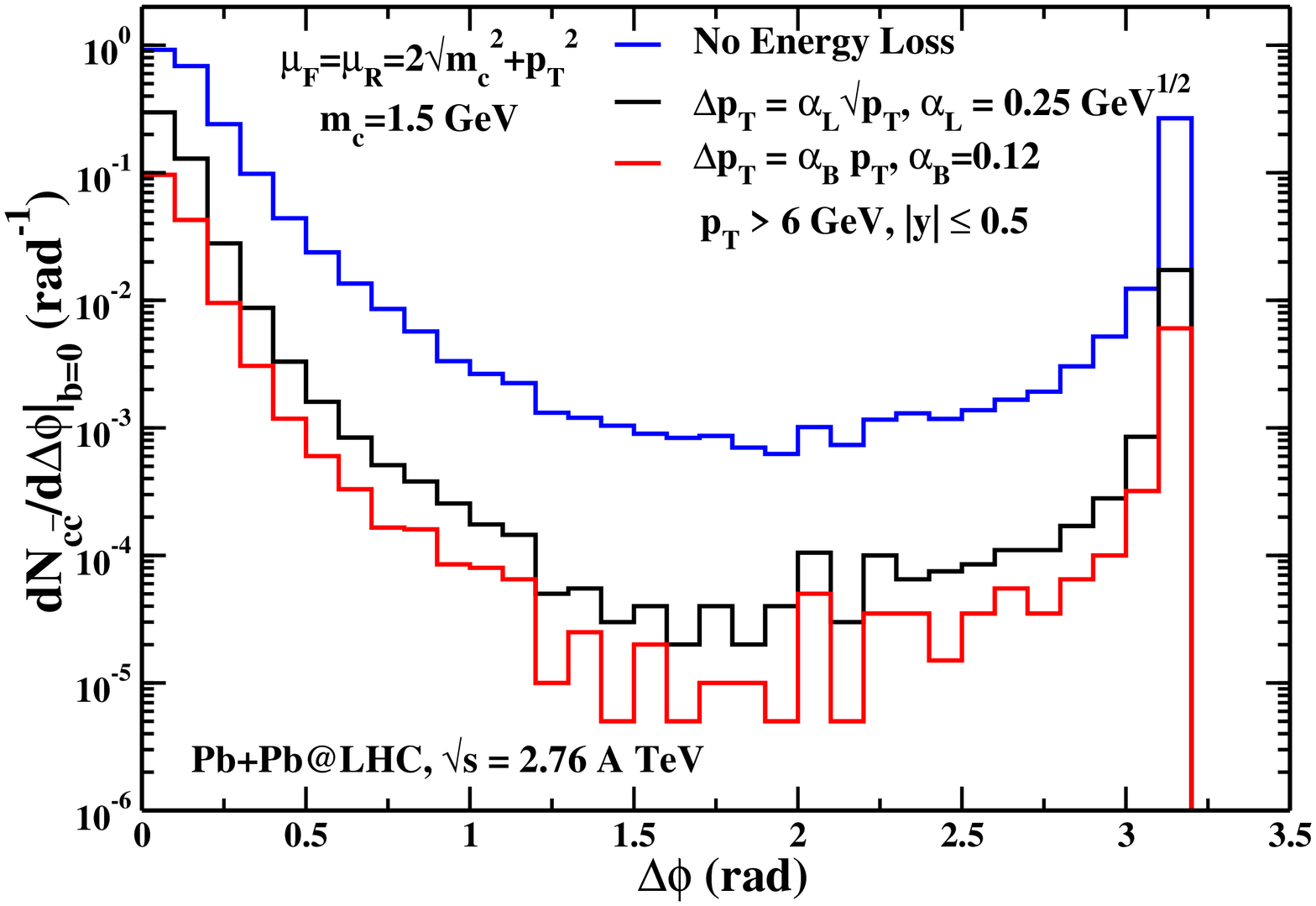}
\caption{(Colour on-line)$dN/d\Delta\phi$ vs $\Delta\phi$ of
$c\overline{c}$ pair for (upper left)$p_T$ $<$ 2.0 GeV, (upper right)$p_T$ $>$ 2.0 GeV.
(lower left)$p_T$ $<$ 6.0 GeV, (lower right)$p_T$ $>$ 6.0 GeV.}
\label{fig2}
\end{center}
\end{figure*}

\begin{figure*}[h]
\begin{center}
\includegraphics[height=3in,width=2.8in,angle=270]{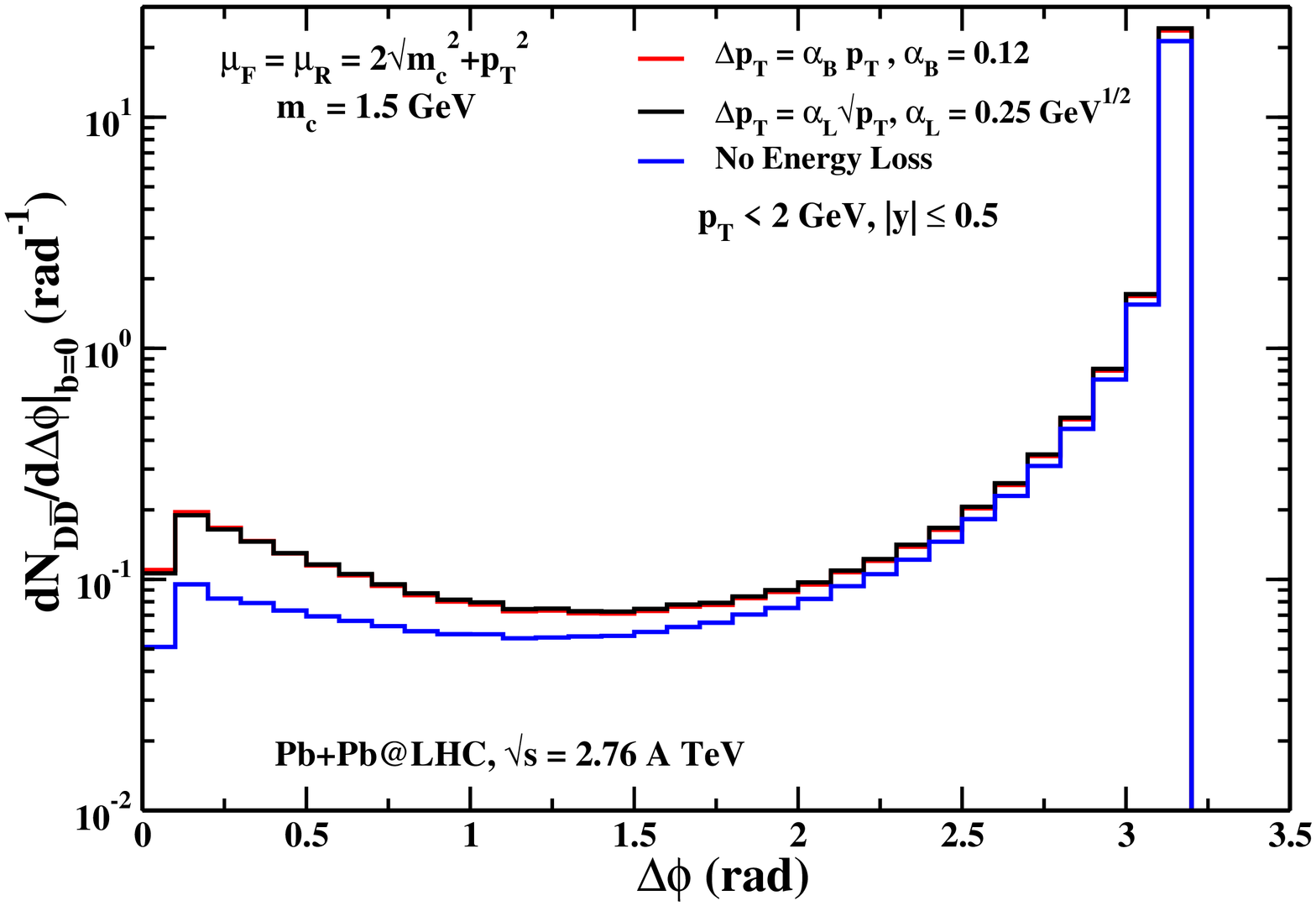}
\includegraphics[height=3in,width=2.8in,angle=270]{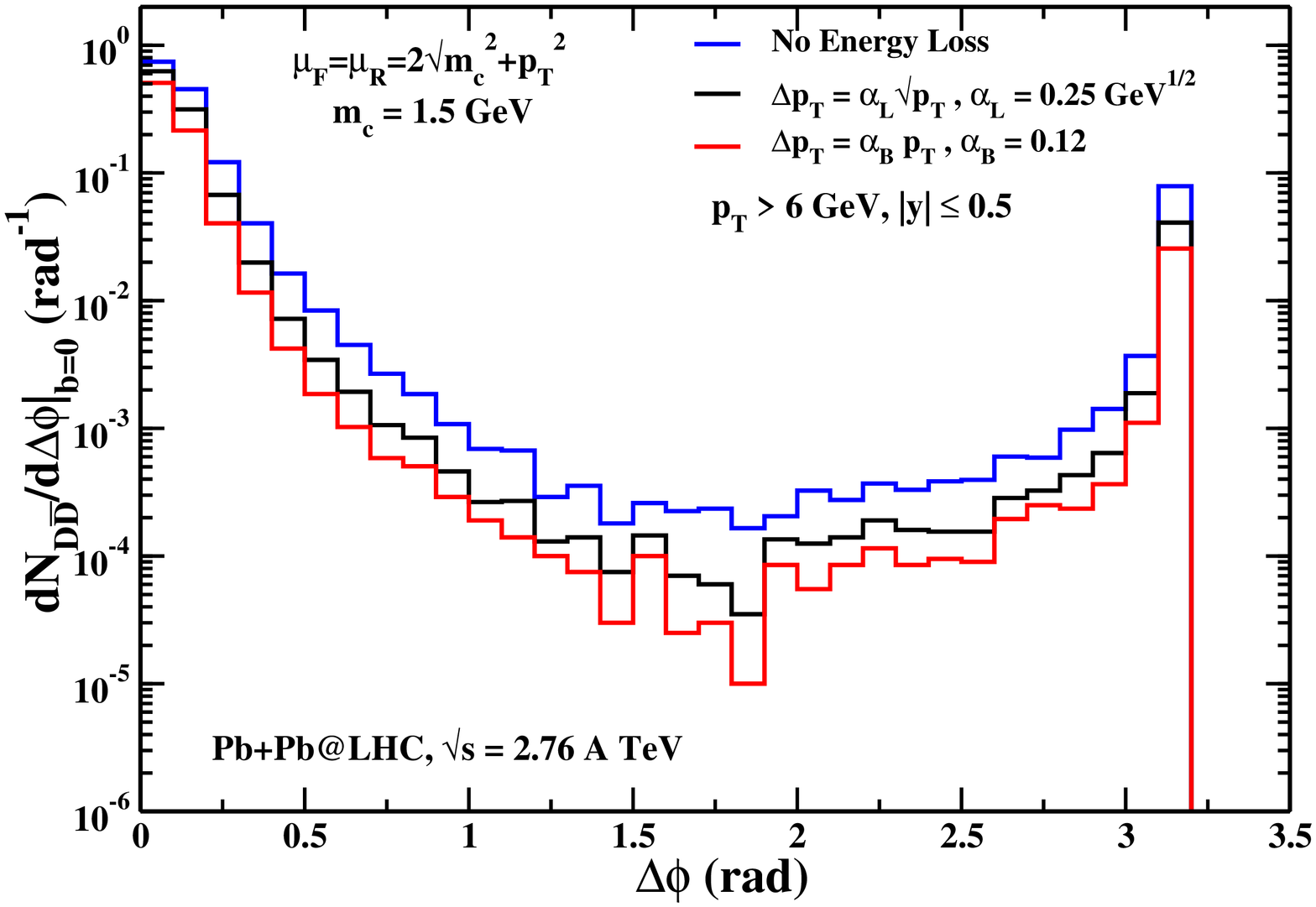}
\caption{(Colour on-line)same as Fig.2, $dN/d\Delta\phi$ vs $\Delta\phi$ of
$D\overline{D}$ pair for (left)$p_T$ $<$ 2.0 GeV, (right)$p_T$ $>$ 6.0 GeV.}
\label{fig3}
\end{center}
\end{figure*}

\begin{figure*}[h]
\begin{center}
\includegraphics[height=3in,width=3in,angle=270]{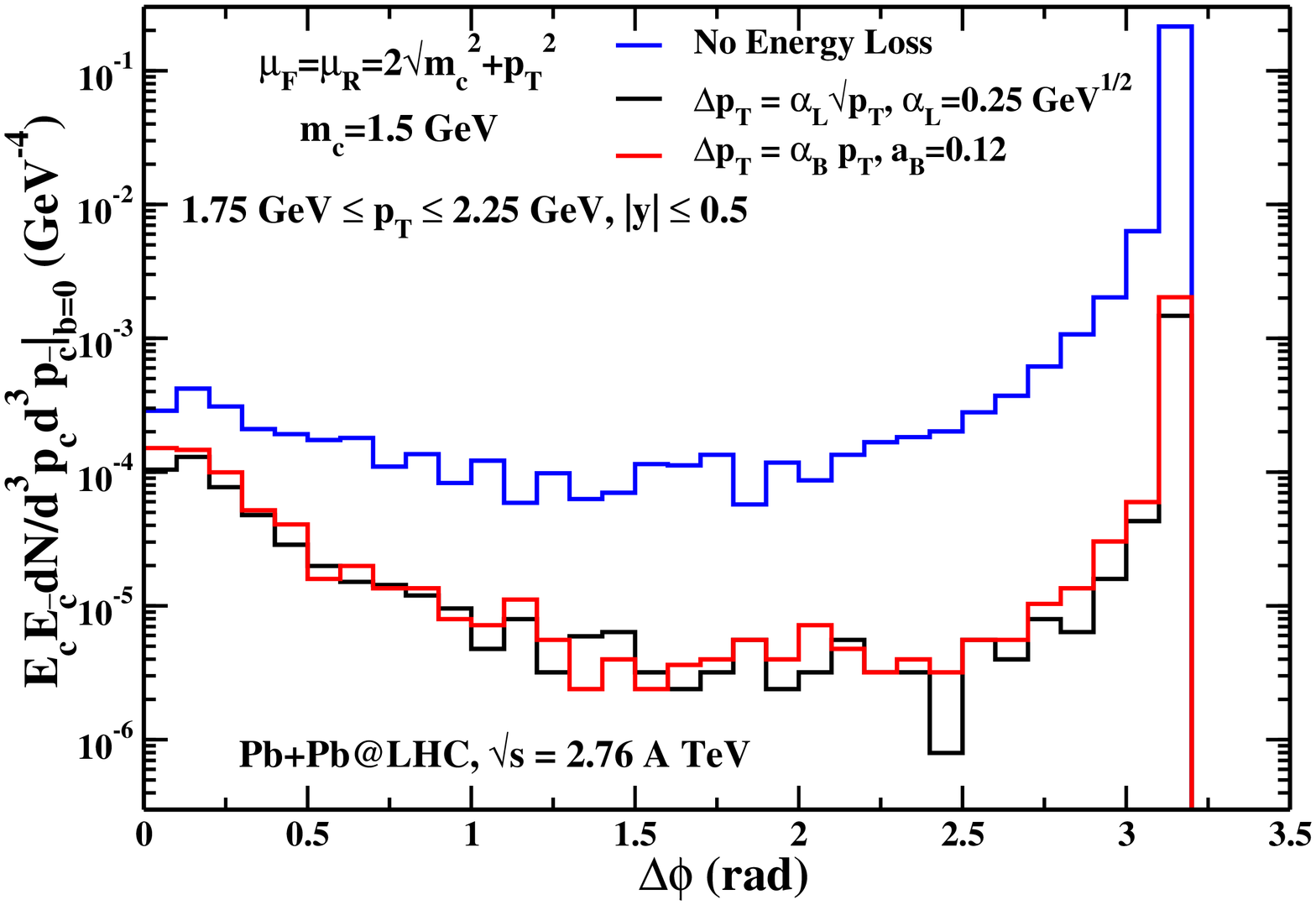}
\includegraphics[height=3in,width=3in,angle=270]{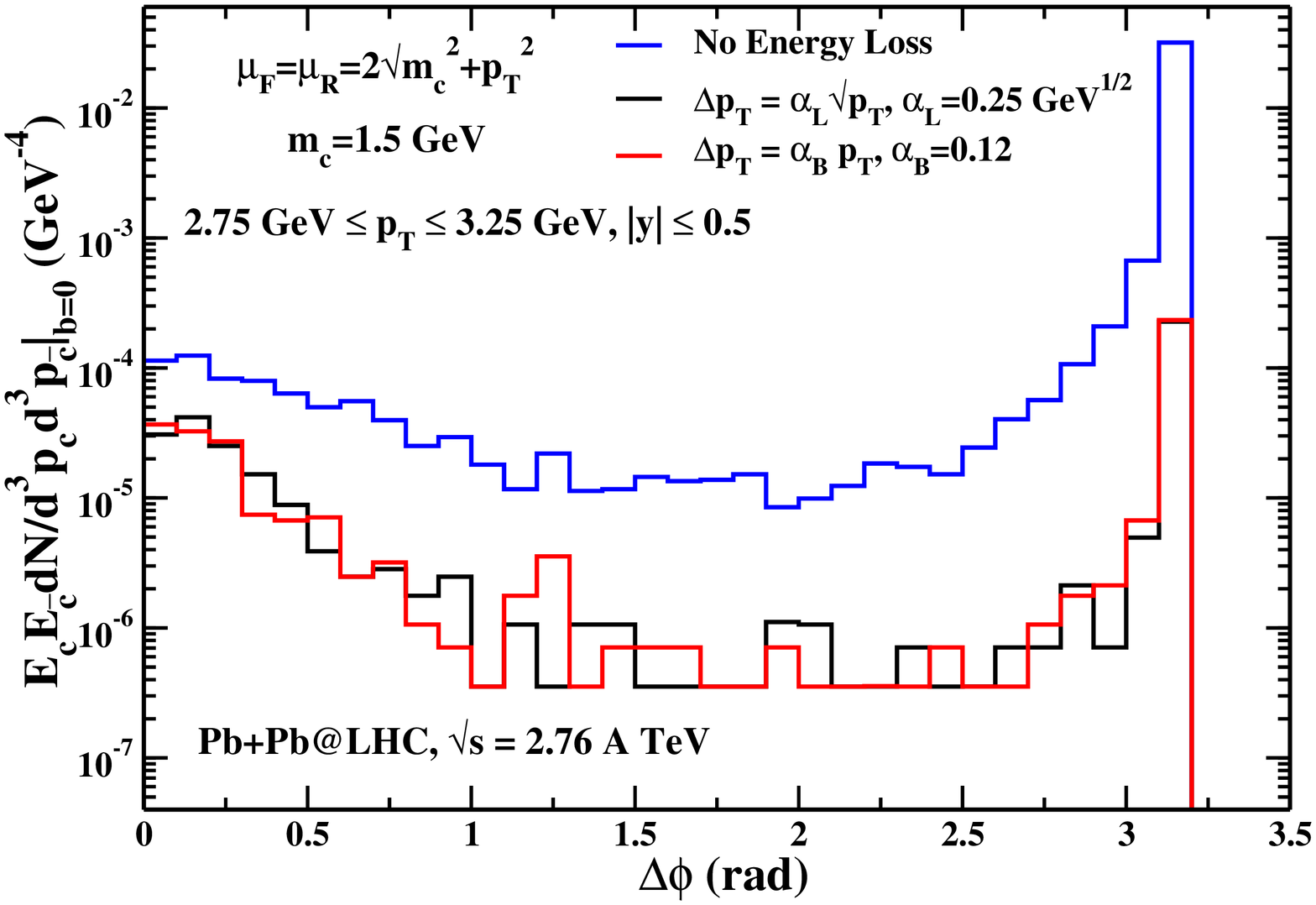}
\caption{(Colour on-line)Azimuhtal correlation of $c\overline{c}$ pair
for (left) $<p_T >$=2.0 GeV, (right) $<p_T >$=3.0 GeV}
\label{fig4}
\end{center}
\end{figure*}

\begin{figure*}[h]
\begin{center}
\includegraphics[height=6in,width=4in,angle=270]{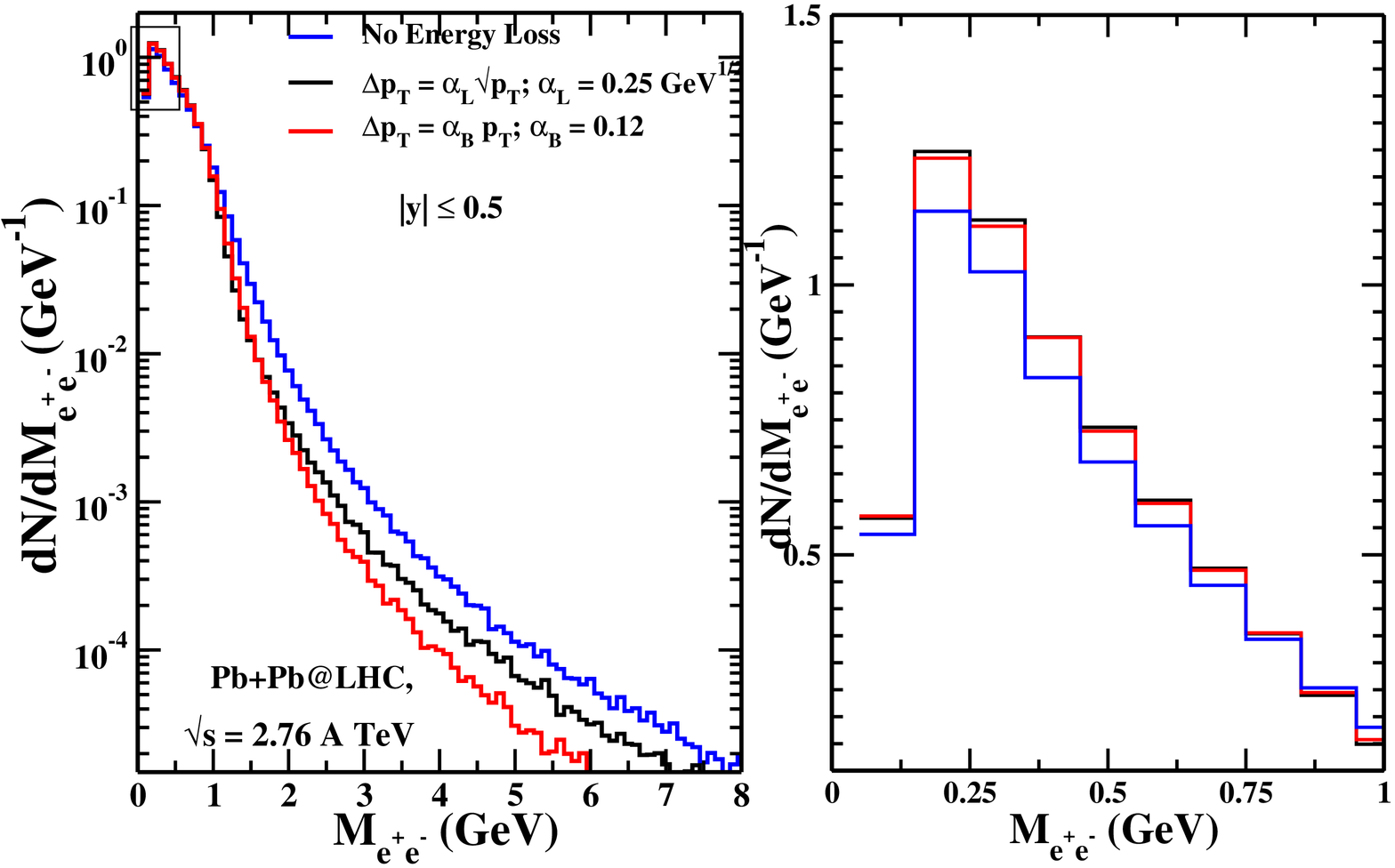}
\caption{(Colour on-line)Invariant mass distribution for di-electron
(inset)Increase in di-electron spectrum for $M_{e^{+}e^{-}} <$1.0 GeV, shown in linear scale. }
\label{fig5}
\end{center}
\end{figure*}

\section{Results and Discussions}
We have used NLO-MNR code ~\cite{nlomnr} with CTEQ5M structure function 
for estimating charm cross-section
for all leading and next-to-leading pQCD processes. The
scaling factor used is 2$\sqrt{m_{c}^{2}+p_{T}^{2}}$ with
$m_{c}$=1.5 GeV. In this paper we have used two
different values for parameter '$\beta$'=1.0 for B-H type and
$\beta$=0.5 for LPM type of energy loss mechanisms respectively.
Correspondingly $\alpha$ = 0.12 for B-H type
and 0.25 GeV$^{1/2}$ for LPM type are taken as the best values at
$\sqrt{s}$ = 2760 GeV/nucleon. The
entire calculation is done for central collision (b=0fm) and
for mid rapidity, -0.5$\leq$y$\leq$0.5

To check the consistency in our results we use two different partonic
structure functions one of which is CTEQ5M and other an old one MRS125.
The comparison is shown in Fig. \ref{fig1}, where the difference in the
two distributions is very small and the shape almost identical. However
more recent structure functions like CTEQ6M and CTEQ6.6 etc. must be
used in order to have more up-to-date results. These issues will
be addressed in our next publication on heavy quark correlation.

Next let us recall that LO contribution can be
differentiated from NLO contribution with different $p_T$
cuts on charm momentum.
Leading order processes give back
to back charm pairs which are entirely visible around $\Delta\phi$=$\pi$
while NLO contribution is distributed from $\Delta\phi$=0 -- $\pi$.

In Fig. \ref{fig2}, we show our results for $dN_{c\bar{c}}/d\phi$
for different $p_T$ cuts. Realizing that all heavy quarks
now appear with reduced momenta, we see that if we look
at $p_T$ $<$ 2 GeV or $p_T$ $<$ 6 GeV, then the back-to-back
correlation rise by up to a factor of 10 for $\phi$ = 0.
The results for $p_T$ $>$ 2 GeV or $p_T$ $>$ 6 GeV are
more dramatic in that the $\phi$=$\pi$
correlation now reduces by more than a factor of 10
while that for $\phi$=0 decreases from its value for no energy loss.

We show $dN_{D\bar{D}}/d\Delta\phi$ for $p_T$ $>$ 6.0 GeV and
$p_T$ $<$ 2.0 GeV in Fig. \ref{fig3}. Comparing it with Fig. \ref{fig2} for same $p_T$ regions, we observe
certain differences which we now discuss. For $p_T$ $<$ 2.0 GeV, we observe
that D meson distribution is slightly higher than charm spectrum at $\Delta\phi$ = $\pi$,
although the order of magnitude remains same. While at $\Delta\phi$ = 0, the situation is reversed.
Similar observations are noted when figures at $p_T$ $>$ 6.0 GeV are compared.
We feel that the above differences are caused by fragmentation
function, D(z), which changes the $p_T$ distribution of charm into $p_T$ distribution of D mesons
with, 0$\leq$z$\leq$1. Thus the correlation spectra of charm and D mesons may 
appear slightly different when we look into particular $p_T$ regions. 
Finally it can be mentioned, D mesons rather than charms are observed in experiments. So
 calculating D meson correlation and comparing it with charm will give us deeper insights into
the correlation study.

In Fig. \ref{fig4}, we have $E_c\,E_{\bar{c}}dN/d^{3}p_c\,d^{3}p_{\bar{c}}$ for charm average $p_T$
of 2 GeV and 3 GeV. The figure shows change in azimuthal correlation of
charm pairs with pairs at $\Delta\phi$=$\pi$ decreased considerably by
inclusion of the energy loss mechanism.

To discuss our simple model of charm quark energy loss,
we find that most of the charm pairs
not only lose energy to shift to the lower
momentum region but also back-to-back correlation for
many charm pair is altered to almost collinear pairs.
Also we find that two different energy loss mechanisms
included in our study do not give much different outcomes.
Further investigating at much higher momentum regions might bring
out the differences between various energy loss mechanisms.
The correlation study can be enriched if expanding
medium is included in addition to energy loss by charms.

Next we move to our results for correlated decay of charm.
In Fig. \ref{fig5}, we have $dN/dM_{e^{+}e^{-}}$ for di-electrons from correlated charm decay.
We can recall that there is enhancement in D mesons as well single non-photonic
electrons due to the effects of large drag on charm quark moving through QGP.
Here we find a similar enhancement in di-electron spectrum at midrapidity.
For $M_{e^{+}e^{-}}$ less than 1 GeV, there is increase in $dN/dM_{e^{+}e^{-}}$
by almost 12\% which is quite noteworthy considering our model to be simple
empirical mechanism of energy loss.

\section{Summary}
We have studied correlation of charm, D mesons as well as correlated decay of charm
using NLO-pQCD processes and a simple empirical model for energy loss.
The azimuthal correlations of charm show change when energy loss
mechanisms are implemented along with cuts on charm transverse momentum. 
In case of di-electron distribution,
energy loss enhances the electron spectrum slightly for low invariant mass.

\section*{Acknowledgments}
One of us (MY) acknowledges financial support of the
Department of Atomic Energy, Government of India
during the course of these studies.


\section*{References}
\begin{thebibliography}{110}
\bibitem{largedrag} G.~D.~Moore and D.~Teaney, Phys. \ Rev. \ {\bf C71}, 064904 (2005);
                     S.~Cao, G.~-Y.~Qin and S.~A.~Bass, arXiv:1205.2396 [nucl-th]
                     and arXiv:1209.5410 [nucl-th];
                     M.~Younus and D.~K.~Srivastava, J.\ Phys.\ {\bf G39}, 095003 (2012).

\bibitem{raa} A.~Adare et al (PHENIX Collaboration) Phys. \ Rev. \ Lett. \ {\bf 98} 172301 (2007);
              B.~I.~Abelev et al Phys. \ Rev. \ Lett. \ {\bf 98} 192301 (2007);
              B.~Abelev et al (ALICE Collaboration) arxiv (2012)

\bibitem{deadcone} Y.~L.~Dokhshitzer, V.~A.~Khoze and S.~I.~Troian, J. \ Phys. \ {\bf G17}, 1602 (1991);
                   Y.~L.~Dokhshitzer, and D.~E.~Kharzeev, Phys. \ Lett. \ {\bf B519}, 199 (2001)
                   R.~Thomas, B.~Kampfer and G.~Soff, Acta. \ Phys. \ Hung. \ {\bf A22} 83 (2005)

\bibitem{dedx} R.~Abir, U.~Jamil and D.~K.~Srivastava,
               Phys. \ lett. \ {\bf B715}, 183 (2012).

\bibitem{corr} M.~Younus and D.~K.~Srivastava, J. \ Phys. \ {\bf G39} 025001, (2012)

\bibitem{flow} N.~Xu, X.~Zhu and P.~Zhuang, Phys. \ Rev. \ Lett. {\bf 100}, 152301 (2008);
               N.~Xu, X.~Zhu and P.~Zhuang, J. \ Phys. \ {\bf G36}, 064025 (2009).

\bibitem{cronin} G.~G.~Barnafoldi, P.~Levai, G.~Fai, G.~Papp and
                  B.~A.~Cole, Int. \ J. \ Mod. \ Phys. \ {\bf E16} 1927, (2007);
                  M.~He, R.~J.~Fries and R.~Rapp (arXiv: 1204.4442 [nucl-th]).

\bibitem{kampfer} B.~Kampfer, O.~P.~Pavlenko and K.~Gallmeister,
                  Phys. \ Lett. \ {\bf B419}, 412 (1998).

\bibitem{jamil} E.~Eichten, I.~Hinchliffe, K.~Lane and C.~Quigg,
                Rev. \ Mod. \ Phys. \ {\bf 56}, 4 (1984);
                 U.~Jamil and D.~K.~Srivastava, J. \ Phys. \ {\bf G}:
                 Nucl. \ Part. \ Phys. \ {\bf 37} 085106 (2010)

\bibitem{invM} B.~L.~Combridge, Nucl. \ Phys. \ {\bf B151}, 429 (1979).

\bibitem{empeloss} X.~N.~Wang, Z.~Huang and I.~Sarcevic Phys. \ Rev. \ Lett. \ {\bf 77}, 231 (1996),
                   X.~N.~Wang and Z.~Huang, Phys. \ Rev. \ {\bf C55}, 3047 (1997);
                   S.~De and D.~K.~Srivastava, J. \ Phys. \ {\bf G39} 015001 (2012);
                   B.~Muller Phys. \ Rev. \ {\bf C67} 061901(R) (2003).

\bibitem{peterson} C.~Peterson, D.~Schlatter, I.~Schmitt and P.~M.~Zerwas, Phys. \ Rev. \ {\bf D27}, 105 (1983).

\bibitem{altarelli} G.~Altarelli, N.~Cabibbo, G.~Corbo, L.~Maiani and G.~Martinelli, Nucl. \ Phys. \ {\bf B208}, 365 (1982).

\bibitem{nlomnr} M.~L.~Mangano, P.~Nason and G.~Ridolfi, Nucl. \ Phys. \ {\bf B373}, 295 (1992).

\end{thebibliography}
\end{document}